%% file: manuscript.tex
\documentclass[conference,a4paper]{IEEEtran}

\usepackage{amsmath,graphicx, booktabs, multirow, amssymb, anyfontsize, glossaries, soul}

\input{glossary}

\title{Statistical analysis of inter coding in VVC Test Model (VTM)}

\iftrue
\author{
\IEEEauthorblockN{Yiqun Liu\IEEEauthorrefmark{1}\IEEEauthorrefmark{2}, Mohsen Abdoli\IEEEauthorrefmark{1}, Thomas Guionnet\IEEEauthorrefmark{1}, Christine Guillemot\IEEEauthorrefmark{2}, and Aline Roumy\IEEEauthorrefmark{2}
}

\IEEEauthorblockA
{
\IEEEauthorrefmark{1}Ateme, Rennes, France
}

\IEEEauthorblockA
{
\IEEEauthorrefmark{2}INRIA, Rennes, France
}

}
\fi

\begin{document}%
\maketitle
\begin{abstract}
The promising improvement in compression efficiency of \gls{vvc} compared to \gls{hevc} \cite{sullivan2012overview} comes at the cost of a non-negligible encoder side complexity. The largely increased complexity overhead is a possible obstacle towards its industrial implementation. Many papers have proposed acceleration methods for \gls{vvc}. Still, a better understanding of \gls{vvc} complexity, especially related to new partitions and coding tools, is desirable to help the design of new and better acceleration methods. For this purpose, statistical analyses have been conducted, with a focus on \gls{cu} sizes and inter coding modes.
\end{abstract}
\section{Introduction}
\label{sec:intro}
Standardization of \gls{vvc} in 2020 has brought significant improvement to the capacity of video compression in terms of bitrate saving. It offers the 50\% compression efficiency \cite{bross2021overview} compared to one of the most efficient video compression standards \gls{hevc}. This improvement in \gls{vvc} is mainly due to newly adopted coding techniques. Most decoder devices could afford the additional complexity brought by these novel coding techniques taking into account current hardware capacities of these devices. Specifically, studies in \cite{pakdaman2020complexity} have shown that the relative complexity of the decoder of \gls{vvc} is from 150\% to 200\% compared to \gls{hevc} in different configurations.

On the contrary, it is far from affordable for real time application for \gls{vvc} encoder side since industrial encoding applications have strict limitations in terms of resources and execution time. Tests in \cite{mercat2021comparative} in \gls{vvc} reference software \gls{vtm} 7 show that the encoding time of \gls{vvc} is 5x, 7x, and 37x times of the encoding of \gls{hevc} in the configurations \gls{ld}, \gls{ra}, and \gls{ai}, respectively. Hence, it is vital to develop acceleration algorithms or methods to largely reduce the encoding complexity while preserving the majority of encoding efficiency. Complexity analysis papers could help researchers to have a clear understanding of what is happening inside a \gls{vvc} encoder (\textit{e.g.} \gls{vtm}), and what potentially interests them for their design process of the acceleration method.

Various studies have contributed to the complexity analysis of \gls{vvc}. In \cite{saldanha2020complexity}, a detailed complexity analysis based on \gls{vvc} intra prediction tools has been performed. Pakdaman \textit{et al.} in \cite{pakdaman2020complexity} have broken down the encoding process into encoding modules such as motion estimation, intra prediction, entropy coding, etc. and then analyzed the complexity partition of modules in multiple encoding configurations. \cite{bossen2021vvc} reviews complexity aspects of the different modules of the \gls{vvc} standard and provide a complexity breakdown of these modules in a more precise way. In \cite{mercat2021comparative}, \gls{vvc} and \gls{hevc} are compared in terms of rate-distortion and complexity analysis. These aforementioned papers present complexity analysis at the level of encoding modules for inter coding. Our paper is the first to provide an analysis from \gls{cu} sizes and coding modes perspective for inter coding in \gls{vvc}.

\begin{figure}[!h]
    \centering
    \includegraphics[scale=0.13]{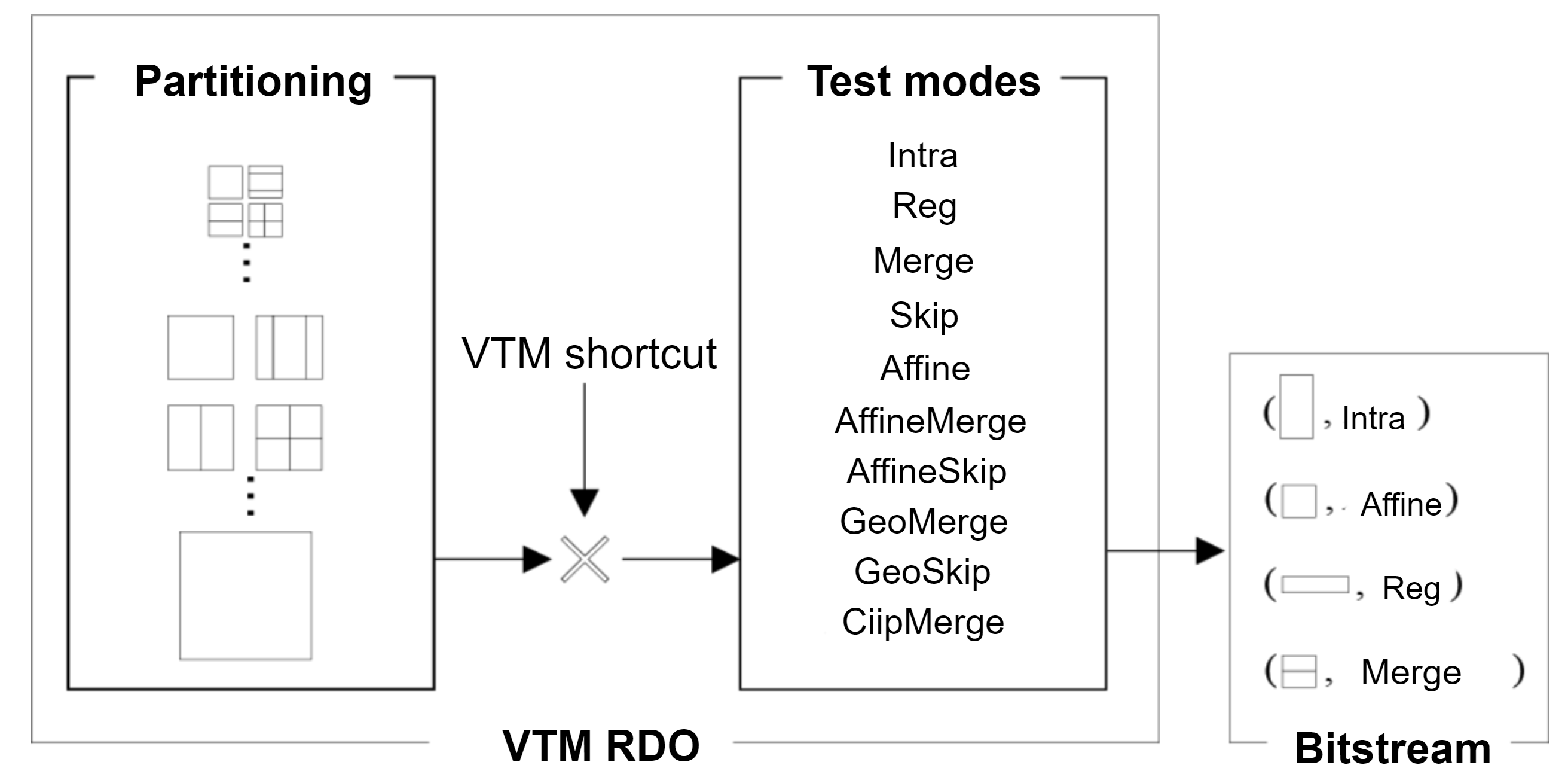}
    \caption{High-level view of the RDO process involving partitioning, test modes and possible \gls{vtm} shortcuts.}
    \label{fig:rdo_high_level}
\end{figure}

In this paper, a statistical analysis of the \gls{rdo} process in inter coding of \gls{vtm}-15.0 is presented. The main focus is put on the statistics of two factors: \gls{cu} sizes and inter coding modes. The goal is to provide useful information for related works aimed at speeding up inter coding in \gls{vvc}. The rest of this paper is organized as follows. Section \ref{sec:rdo} presents a summary of \gls{vvc} specification in terms of \gls{cu} sizes and available coding modes. In Section \ref{sec:statistics}, all statistical observations are presented, which are later analyzed and concluded in Section \ref{sec:analyse}.


\section{RDO of inter coding in VTM}
\label{sec:rdo}

To represent the \gls{rdo} process in \gls{vtm}, there exist numerous coding parameters such as \gls{ipm} of intra coding, \gls{mv} representation mode, choice of transform \textit{etc.}. However, if we ignore these trivial parameters, the \gls{rdo} process could be described as the
search for the best trade-off between bit rate and distortion. More precisely, this search is executed on different coding modes of different \gls{cu} sizes. Therefore, in this paper, the statistics of \gls{cu} sizes and coding modes are jointly considered.

As presented in Fig.\ref{fig:rdo_high_level}, various \gls{cu} sizes are the result of partitioning in the \gls{rdo} process. The partitioning consists of splitting the \gls{cu} of size 128$\times$128 recursively by five split modes, namely \gls{qt} split, \gls{hbt} split, \gls{vbt} split, \gls{htt} split, \gls{vtt} split. Compared to codec \gls{hevc} in which only \gls{qt} is available for partitioning, the added directional splits give rise to a larger variety of \gls{cu} sizes. In \gls{vvc} \gls{cu} size is authorised if its widths and heights are any power of two between 4 and 128, except for sizes 128$\times$4, 128$\times$8, 128$\times$16 and 128$\times$32. It is worth noting that the same \gls{cu} size could be obtained by different series of split modes.

For each \gls{cu}, 12 coding modes are available. Two of these modes, namely hash-based inter prediction and palette modes are not enabled in JVET \gls{ctc}. Hence, they are not included in the analysis of this paper. In the inter prediction of \gls{vvc}, motion compensation is executed after motion estimation. Subsequently, the residuals and \gls{mv} information need to be transmitted. Depending on the transmission of \gls{mv} and residuals, three coding modes are available: \gls{amvp}\cite{chien2021motion} mode, \emph{Merge} mode, and \emph{Skip} mode. For the purpose of simplification, we refer to \gls{amvp} as \emph{Reg} mode for the remainder of the paper. Before transmitting \gls{mv}s, a candidate list of \gls{mv}s is constructed based on the spatial
and temporal neighboring \gls{cu}s by exploiting the correlations of \gls{mv}s between them. Four types of inter prediction data can be signaled, including the index of reference frame (\textit{i.e.} Ref Frame Idx), the index of the best \gls{mv} candidate (\textit{i.e.} \gls{mv} Cand Idx), the difference between the best candidate and \gls{mv} determined by motion estimation (\textit{i.e.} \gls{mvd}), and residuals. Tab.\ref{tab:motion_data_coding} presents the signaled data types for \emph{Reg}, \emph{Merge} and \emph{Skip}.

\begin{table}[!h]
    \centering
    \caption{Data types to transmit for motion data coding}
\fontsize{9pt}{9pt}\selectfont    
    \begin{tabular}{c|c|c|c|c}
    \toprule  \hline
         \multirow{5}{*} & Ref Frame Idx & MV Cand Idx & MVD & Residuals\\ \cline{1-5}
         \emph{Reg} & \checkmark & \checkmark & \checkmark & \checkmark \\\cline{1-5}
         \emph{Merge} & X & \checkmark & X & \checkmark \\\cline{1-5} \emph{Skip} & X & \checkmark & X & X \\\cline{1-5}
    \bottomrule
    \end{tabular}
    \label{tab:motion_data_coding}
\end{table}

In addition to the \emph{Affine} mode and the \emph{Intra} mode, two novel coding modes are available in \gls{vvc}. For \gls{cu}s coded in merge mode, \gls{ciip}\cite{ciip} combines the inter prediction and the intra prediction to form a final prediction. The Geometric Partitioning Mode, denoted as \emph{Geo}, is designed to better predict moving objects in video. Additionally, \emph{Geo} is conventionally coded with \emph{Merge} mode. For inter coding configuration, coding modes of \emph{Reg}, \emph{Merge} and \emph{Skip} could combine
with \emph{Affine}, \emph{CIIP}, and \emph{Geo}, which results in a total of 10 coding modes: \emph{Intra}, \emph{Reg}, \emph{Merge}. \emph{Skip}, \emph{Affine}, \emph{AffineMerge}, \emph{AffineSkip}, \emph{GeoMerge}, \emph{GeoSkip}, \emph{CiipMerge}. Statistics of these coding modes are further collected and analyzed in the following part.

Although \gls{vvc} is computationally expensive, the \gls{jvet} group has already adopted various shortcuts or conditional early exits as presented in \cite{wieckowski2019fast} for the \gls{vtm}. We have deactivated existing shortcuts in \gls{vtm}-15.0 and evaluated its performance. As a result, the complexity increases by 138\%. Furthermore, the performance of the tested encoder (\textit{i.e.} without shortcuts) is 0.76\% better than the reference encoder (\textit{i.e.} with shortcuts), in terms of BD-rate. This trade-off might be interpreted as an indicator that the shortcuts in \gls{vtm} are efficient in terms of identifying useless tests and partitioning depths. Many of shortcuts are based on history of the tested split modes. However, aspects of \gls{cu} sizes and coding modes are overlooked.

\begin{figure}[!h]
    \centering
    \includegraphics[width=0.8\linewidth]{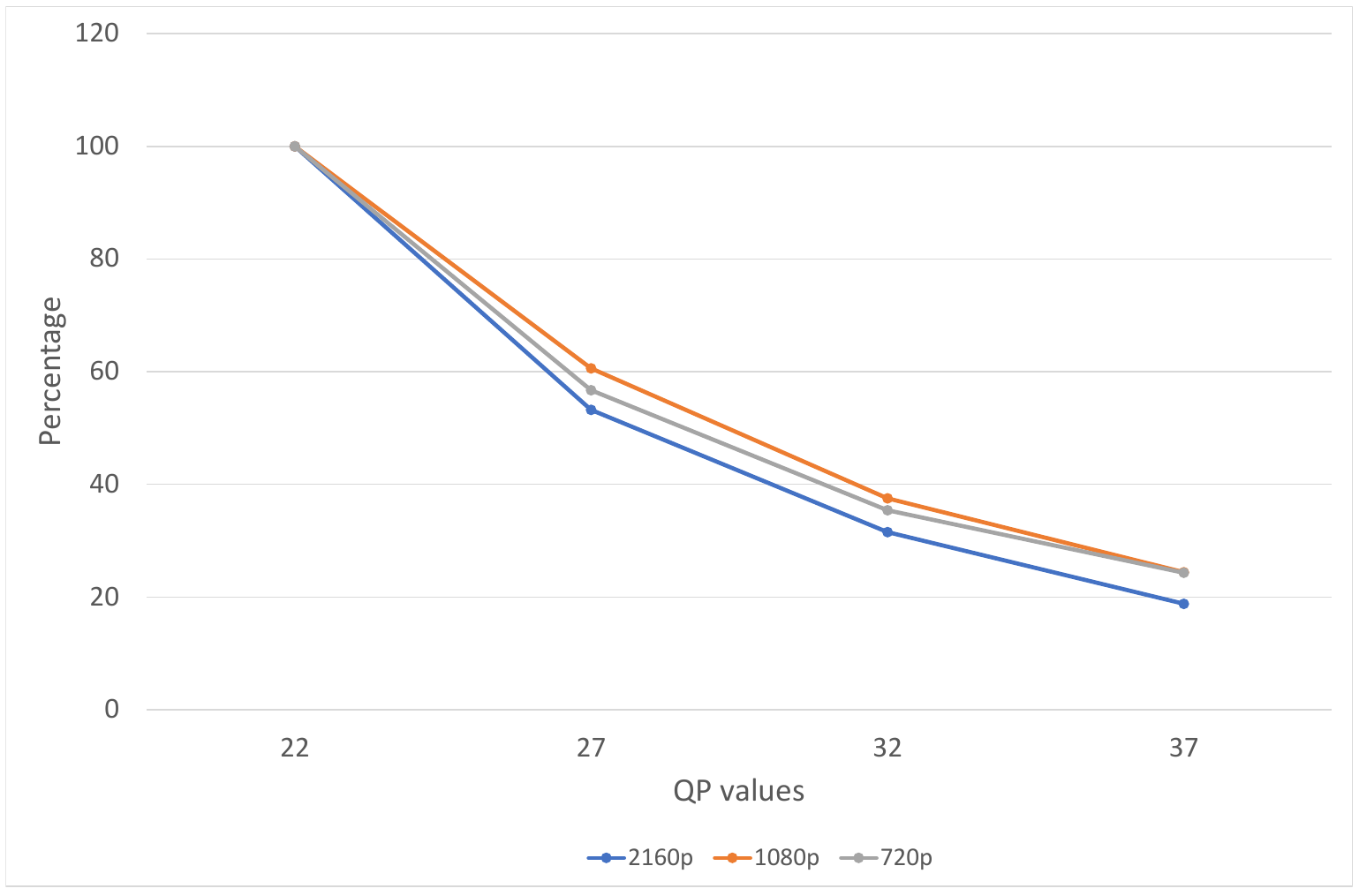}
    \caption{Encoding time in different QPs comparing to QP 22} 
    \label{fig:time_vs_qp} 
\end{figure}




\section{Statistics}
\label{sec:statistics}

Our main purpose in the following analysis is to find \gls{cu} sizes and/or coding modes with a relatively high complexity occupation and a low selection rate in the \gls{rdo} process. From the perspective of encoder acceleration, \gls{cu} sizes or coding modes with a higher complexity portion and a significantly lower selection rate are more favorable to the design of acceleration rules based on \gls{cu} size/coding mode. The size or mode with larger complexity portion has more potential in accelerating. Lower selection rate indicates it is less likely to make wrong decisions when skipping the \gls{rdo} of current \gls{cu} size or mode.

All our experiments and analyses are performed on the first 64 frames of the \gls{ctc} sequences in the \gls{ragop32} configuration in which intra frames are excluded. Exceptionally, Fig.\ref{fig:time_vs_qp} is based on sequences in Class A, B, E of \gls{ctc}.

\begin{figure}[!h]
    \centering
    \includegraphics[width=1\linewidth]{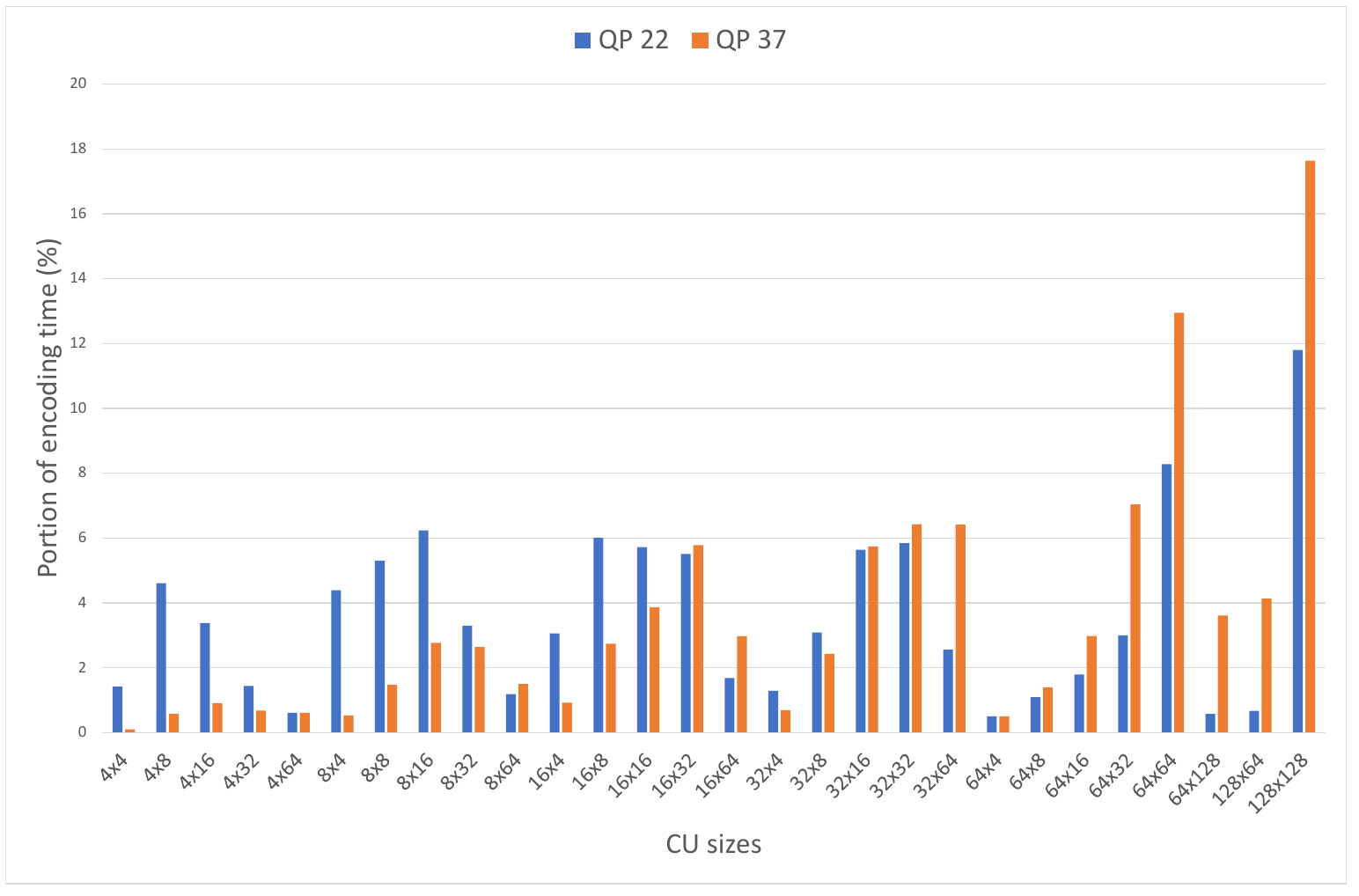}
    \caption{Complexity distribution for \gls{cu} sizes in QP22 and QP37} 
    \label{fig:complexity_vs_qp} 
\end{figure}

The encoding complexity for Chrominance channel only accounts for a small part comparing to Luminance channel. Thus we focus on Luminance channel in the remaining of the paper. From a high-level perspective, the encoding complexity of \gls{vtm} significantly depends on the selected \gls{qp}. Particularly, larger \gls{qp} values tend to have faster encoding with the \gls{vtm}. Fig. \ref{fig:time_vs_qp} is obtained by measuring the encoding times of sequences of resolution 2160p, 1080p and 720p in \gls{qp} 22, 27, 32 and 37. Then the average ratio is calculated between the encoding time of each \gls{qp} and that of \gls{qp} 22 is calculated. It shows that the encoding time at \gls{qp} 22 could be five times as much as at \gls{qp} 37.

Fig.\ref{fig:complexity_vs_qp} shows the percentage distribution of the encoding time spent on different \gls{cu} sizes in \gls{qp} 22 and \gls{qp} 37. In addition to the fact that the overall encoding time is higher for \gls{qp} 22, it can be observed that a relatively higher portion of the time in \gls{qp} 22 is passed on smaller \gls{cu} sizes. This could be partly explained by the existing shortcuts in \gls{vtm} disallowing excessively small \gls{cu}s in \gls{qp} 37. We could declare that larger \gls{cu} sizes are in general more crucial to speeding up the partitioning process, especially \gls{cu} 64$\times$64 and 128$\times$128 which take in total from 20\% complexity in \gls{qp} 22 to 30\% in \gls{qp} 37.

\begin{figure}[!h]
    \centering
    \includegraphics[width=1\linewidth]{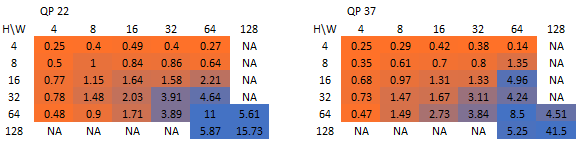}
    \caption{Selection rate for different \gls{cu} sizes} 
    \label{fig:block_prob_test_encode_count_vs_qp} 
\end{figure}

In another test, the selection percentages of different \gls{cu} sizes are calculated. This metric is defined as the ratio between the total number of times it is selected and the total number of times a \gls{cu} size is tested. Fig. \ref{fig:block_prob_test_encode_count_vs_qp} shows the values of this metric in \gls{qp} 22 and 37. As we can see from this figure, larger \gls{cu} sizes correspond to larger selection rates compared with smaller \gls{cu}s. Another phenomenon worth noting is that the selection rate of 128$\times$128 increase dramatically from 14\% in \gls{qp} 22 to 37\% in \gls{qp} 37.

Combining the above figures, it is observed that \gls{cu} sizes such as 16$\times$8, 8$\times$16 and 16$\times$16 are sizes with low selection rate and high complexity. For example, 16$\times$16 \gls{cu}s have the same level of complexity, while its selection rate is half of 32$\times$32 \gls{cu}s in \gls{qp} 22.

\begin{figure}[!h]
    \centering
    \includegraphics[width=1\linewidth]{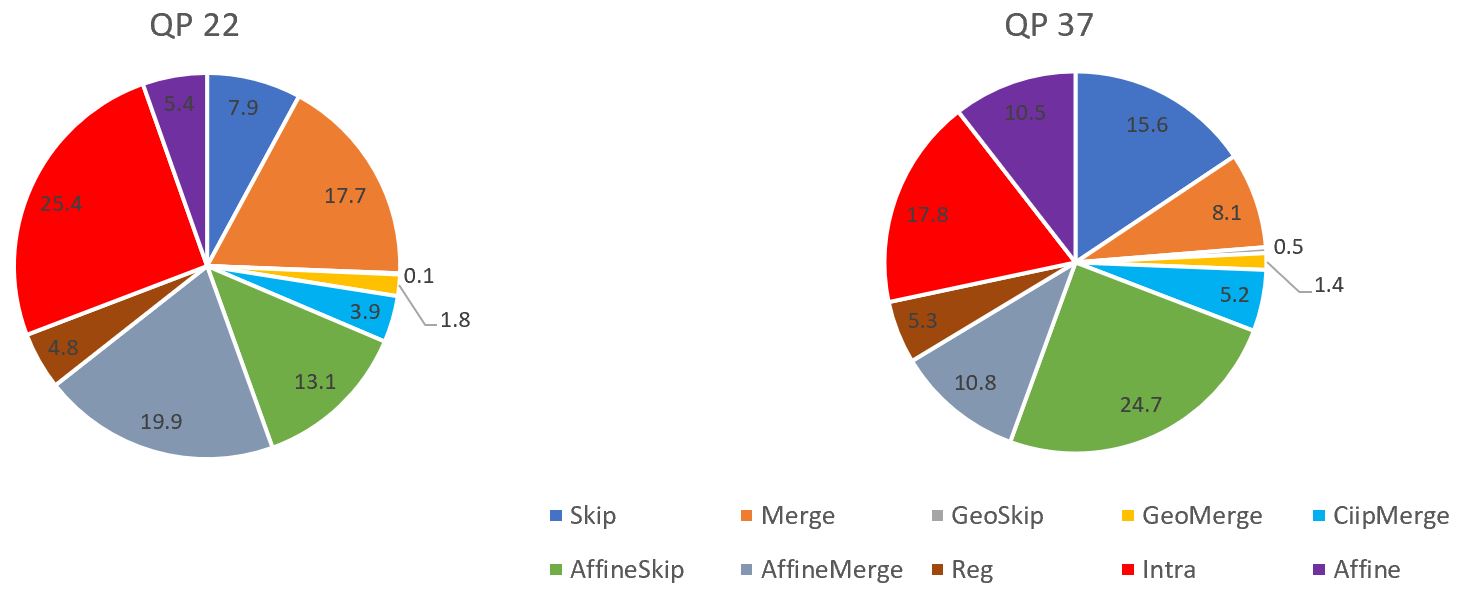}
    \caption{Pie chart of complexities of inter coding modes} 
    \label{fig:test_vs_encode_mode_Q37}
\end{figure}

To take one step further in the statistical analysis of inter coding, we present how different inter coding modes are involved in \gls{rdo} search. The first experiment in Fig.\ref{fig:test_vs_encode_mode_Q37} presents the distribution of the encoding time at inter coding mode level in \gls{qp} 22 and \gls{qp} 37. 

\begin{figure}[!h]
    \centering
    \includegraphics[width=1\linewidth]{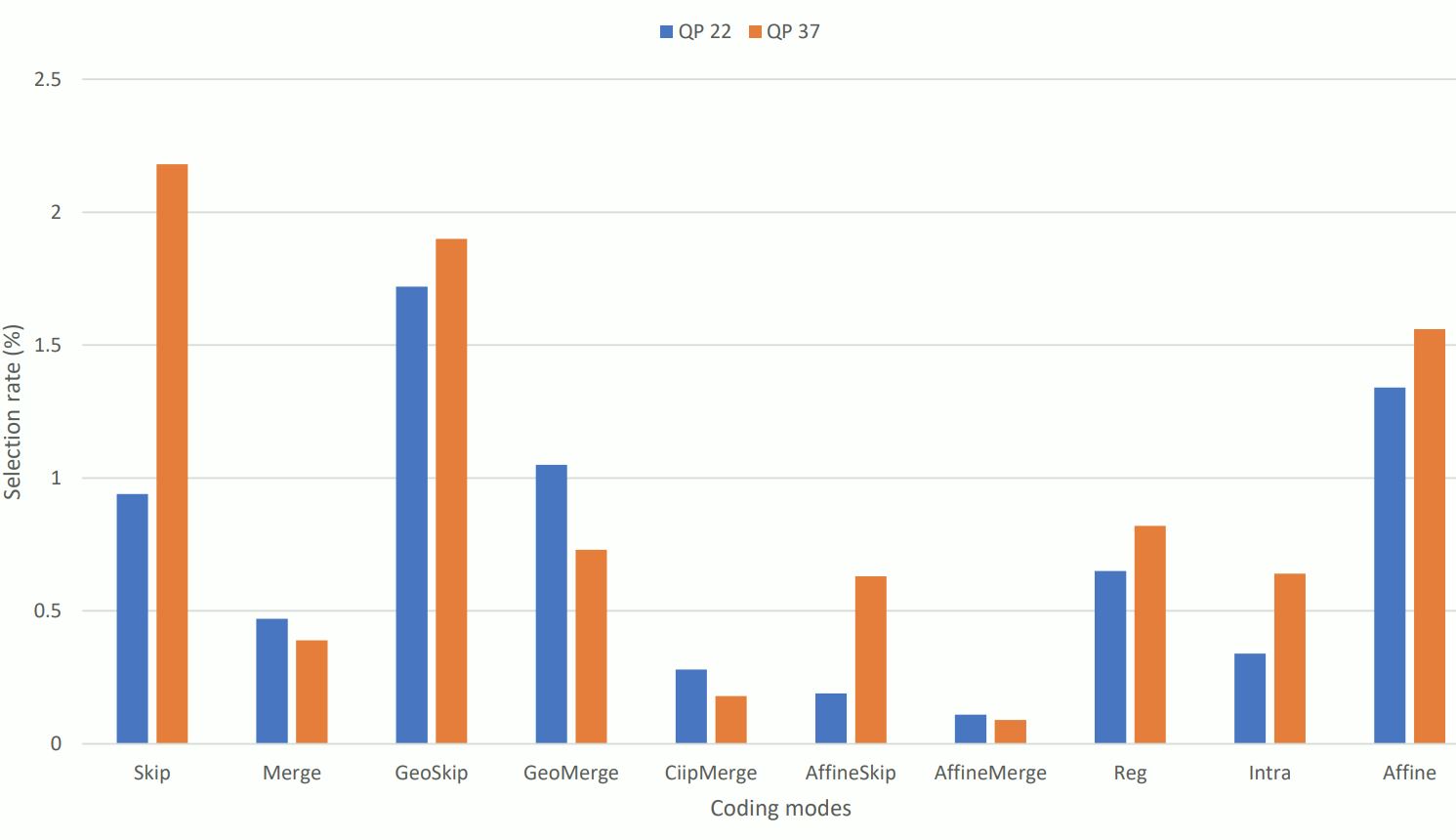}
    \caption{Selection rate of inter coding modes} 
    \label{fig:selection_rate_inter_mode} 
\end{figure}

In general, \emph{Intra}, \emph{AffineMerge}, \emph{AffineSkip}, \emph{Merge}, and \emph{Skip} are main contributors to encoding time. Fig.\ref{fig:selection_rate_inter_mode} provides selection rates as the ratio between the number of selected inter coding modes and number of tested modes. We could observe that the three modes, namely \emph{AffineMerge}, \emph{AffineSkip}, and \emph{Merge} have relatively low selection rates, although they collectively account for nearly half of the complexity.

\begin{figure*}[!h]
    \centering
    \includegraphics[width=1\linewidth]{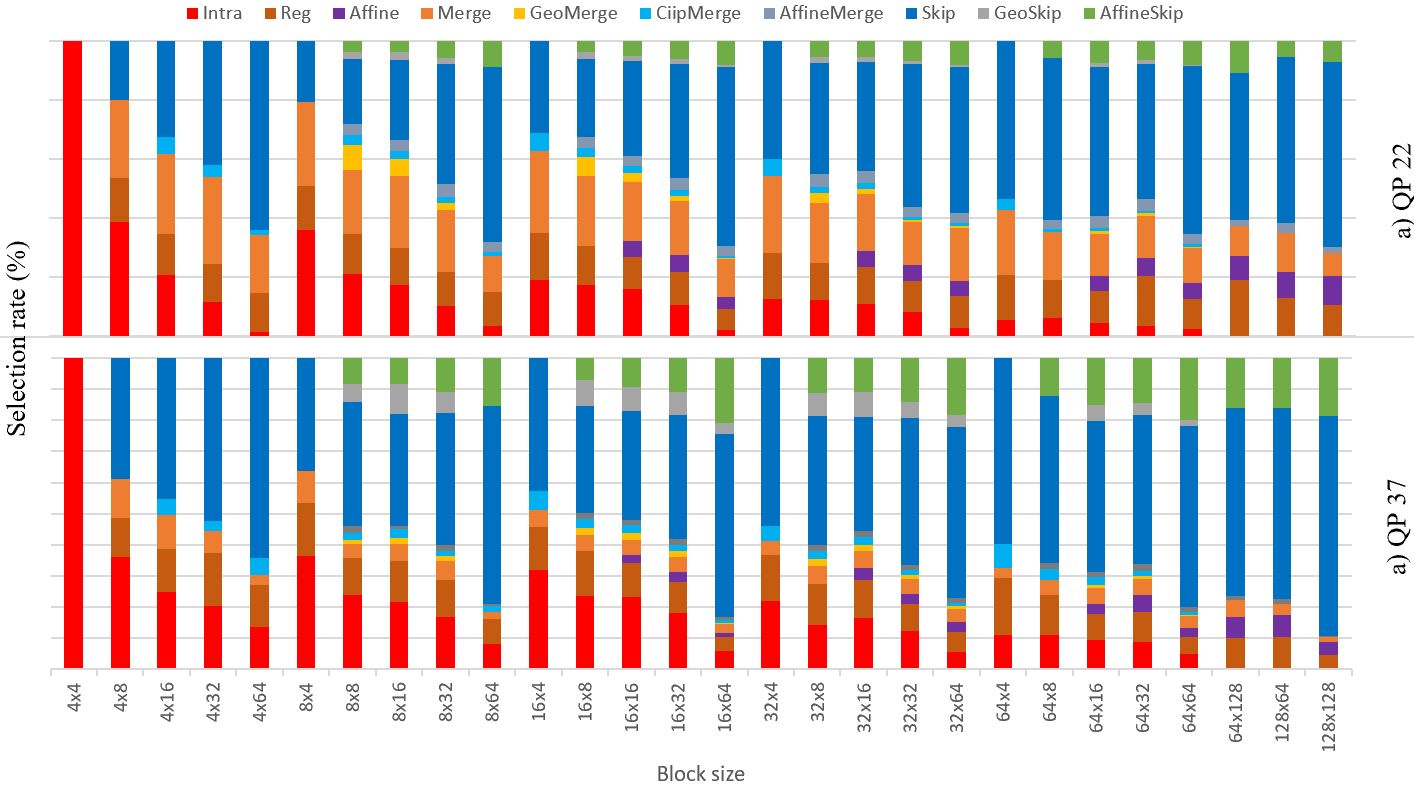}
    \caption{Stacked chart of selected inter modes in different \gls{cu} sizes} 
    \label{fig:mode_encode_rate_block_size} 
\end{figure*}

Fig.\ref{fig:mode_encode_rate_block_size} shows the distribution of inter modes for encoded \gls{cu}s of different sizes. The fact that the aforementioned three coding modes are less chosen could also be proved by this figure. We find that the number of \emph{Skip} is dominant for most \gls{cu} sizes and that the number of these three modes is relatively small, which is consistent with Fig.\ref{fig:selection_rate_inter_mode}. From Fig.\ref{fig:mode_encode_rate_block_size} we also observe that smaller \gls{cu}s tend to be encoded with intra mode. Another remark is that the skip modes (\textit{i.e.} \emph{Skip}, \emph{AffineSkip}, and \emph{GeoSkip}) are more frequently selected for larger \gls{cu}s. It is probably because the residual of larger \gls{cu}s is more expensive to be encoded. In addition, \emph{Merge} mode has a higher chance to be selected in smaller \gls{qp} which is in contrast to \emph{AffineSkip} and \emph{GeoSkip}.
For some \gls{cu} sizes, we could make shortcuts or conditions for early termination for \gls{rdo} of inter modes which are rarely selected to speed up the encoding process, such as the \emph{AffineMerge} mode with merely 1.9\% selected for \gls{cu} 128$\times$128.


\section{Analysis and Conclusion}
\label{sec:analyse}

In this study, complexity analysis of \gls{cu} sizes and inter coding modes has been combined with selection rate analysis. From the perspective of \gls{cu} size, \gls{cu} sizes with high complexity generally correspond to a high selection rate. \gls{cu} sizes 128$\times$128 and 64$\times$64 are responsible for one-third of the complexity. In addition, \gls{cu} sizes such as 16$\times$8, 8$\times$16, 16$\times$16 exhibit relatively low selection rate while requiring a significant share of the overall complexity. Therefore, these \gls{cu} sizes are relevant targets for acceleration algorithms. From a coding mode perspective, \emph{AffineMerge}, \emph{AffineSkip}, and \emph{Merge} tend to be less likely to be selected. Thus, dedicated shortcuts to adaptively skip these coding modes might be promising. Shortcuts on coding modes and partitioning acceleration method are in different scopes. The former focus on reducing number of \gls{cu} for \gls{rdo}. The latter speeds up \gls{rdo} for \gls{cu} of certain sizes. The combination of these two could lead to a larger speed-up of encoding.

\bibliographystyle{IEEEbib}
\bibliography{references}

\end{document}

%% file: glossary.tex
\newacronym{rdo}{RDO}{Rate-Distortion Optimization}
\newacronym{jvet}{JVET}{Joint Video Exploration Team}
\newacronym{vtm10}{VTM10}{VVC Test Model 10}
\newacronym{vtm15}{VTM15}{VVC Test Model 15}
\newacronym{vtm}{VTM}{VVC Test Model}
\newacronym{hevc}{HEVC}{High Efficiency Video Coding}
\newacronym{bdrate}{BD-rate}{Bjontegaard Delta-Rate}
\newacronym{qtmap}{QTdepthMap}{Quad Tree depth map}
\newacronym{cnn}{CNN}{Convolutional Neural Network}
\newacronym{ctc}{CTC}{Common Test Condition}
\newacronym{ragop32}{RAGOP32}{RandomAccess Group Of Picture 32}
\newacronym{vvc}{VVC}{Versatile Video Coding}
\newacronym{hdr}{HDR}{High Definition Range}
\newacronym{vr}{VR}{Virtual Reality}
\newacronym{qtmt}{QTMT}{quadtree with nested multi-type tree}
\newacronym{ctu}{CTU}{Coding Tree Unit}
\newacronym{ns}{NS}{No Split}
\newacronym{cu}{CU}{Coding Unit}
\newacronym{qt}{QT}{Quaternary Tree}
\newacronym{bt}{BT}{Binary Tree}
\newacronym{mt}{MT}{Multi-type Tree}
\newacronym{tt}{TT}{Ternary-type Tree}
\newacronym{hbt}{HBT}{Horizontal Binary Tree}
\newacronym{rdcost}{RDcost}{RateDistorsion Cost}
\newacronym{vbt}{VBT}{Vertical Binary Tree}
\newacronym{vtt}{VTT}{Vertical Tenary Tree}
\newacronym{htt}{HTT}{Horizontal Tenary Tree}
\newacronym{rf}{RF}{Random Forest}
\newacronym{mltcnn}{MLT-CNN}{multi-level tree CNN}
\newacronym{mv}{MV}{Motion Vector}
\newacronym{ipm}{IPM}{Intra Prediction Mode}
\newacronym{ciip}{CIIP}{Combined Intra-Inter Prediction}
\newacronym{ra}{RA}{Random-Access}
\newacronym{ld}{LD}{Low-Delay}
\newacronym{ai}{AI}{All-Intra}
\newacronym{hm}{HM}{HEVC test Model}
\newacronym{mtt}{MTT}{Multi-Type Tree}

\newacronym{isp}{ISP}{Intra SubPartition}
\newacronym{satd}{SATD}{Sum of Absolute Transformed Difference}
\newacronym{jem}{JEM}{Joint Exploration Model}
\newacronym{bcnn}{B-CNN}{Branch Convolutional Neural Network}
\newacronym{mbmpcnn}{MBMP-CNN}{Multi-Branch Multi-Pooling CNN}
\newacronym{qp}{QP}{Quantization Parameter}
\newacronym{cbam}{CBAM}{Convolutional block attention module}
\newacronym{mae}{MAE}{Mean Absolute Error}

\newacronym{ts}{TS}{Time Saving}

\newacronym{amvp}{AMVP}{Advanced Motion Vector Prediction}

\newacronym{mvd}{MVD}{Motion Vector Difference}

%% file: manuscript.bbl
\begin{thebibliography}{1}

\bibitem{sullivan2012overview}
Gary~J Sullivan, Jens-Rainer Ohm, Woo-Jin Han, and Thomas Wiegand,
\newblock ``Overview of the high efficiency video coding (hevc) standard,''
\newblock {\em IEEE Transactions on circuits and systems for video technology},
  vol. 22, no. 12, pp. 1649--1668, 2012.

\bibitem{bross2021overview}
B.~Bross et~al.,
\newblock ``{Overview of the versatile video coding (VVC) standard and its
  applications},''
\newblock {\em IEEE Trans. on Circuits and Systems for Video Technology}, vol.
  31, no. 10, pp. 3736--3764, 2021.

\bibitem{pakdaman2020complexity}
Farhad Pakdaman, Mohammad~Ali Adelimanesh, Moncef Gabbouj, and Mahmoud~Reza
  Hashemi,
\newblock ``Complexity analysis of next-generation vvc encoding and decoding,''
\newblock in {\em 2020 IEEE International Conference on Image Processing
  (ICIP)}. IEEE, 2020, pp. 3134--3138.

\bibitem{mercat2021comparative}
Alexandre Mercat, Arttu M{\"a}kinen, Joose Sainio, Ari Lemmetti, Marko
  Viitanen, and Jarno Vanne,
\newblock ``Comparative rate-distortion-complexity analysis of vvc and hevc
  video codecs,''
\newblock {\em IEEE Access}, vol. 9, pp. 67813--67828, 2021.

\bibitem{saldanha2020complexity}
M{\'a}rio Saldanha, Gustavo Sanchez, C{\'e}sar Marcon, and Luciano Agostini,
\newblock ``Complexity analysis of vvc intra coding,''
\newblock in {\em 2020 IEEE International Conference on Image Processing
  (ICIP)}. IEEE, 2020, pp. 3119--3123.

\bibitem{bossen2021vvc}
Frank Bossen, Karsten S{\"u}hring, Adam Wieckowski, and Shan Liu,
\newblock ``Vvc complexity and software implementation analysis,''
\newblock {\em IEEE Transactions on Circuits and Systems for Video Technology},
  vol. 31, no. 10, pp. 3765--3778, 2021.

\bibitem{chien2021motion}
Wei-Jung Chien, Li~Zhang, Martin Winken, Xiang Li, Ru-Ling Liao, Han Gao,
  Chih-Wei Hsu, Hongbin Liu, and Chun-Chi Chen,
\newblock ``Motion vector coding and block merging in the versatile video
  coding standard,''
\newblock {\em IEEE Transactions on Circuits and Systems for Video Technology},
  vol. 31, no. 10, pp. 3848--3861, 2021.

\bibitem{ciip}
Xin Jin, King Ngan, and Guangxi Zhu,
\newblock ``Combined inter-intra prediction for high definition video coding,''
\newblock 01 2007.

\bibitem{wieckowski2019fast}
Adam Wieckowski, Jackie Ma, Heiko Schwarz, Detlev Marpe, and Thomas Wiegand,
\newblock ``Fast partitioning decision strategies for the upcoming versatile
  video coding (vvc) standard,''
\newblock in {\em 2019 IEEE International Conference on Image Processing
  (ICIP)}. IEEE, 2019, pp. 4130--4134.

\end{thebibliography}
